# Interlayer electronic hybridization leads to exceptional thickness-dependent vibrational properties in few-layer black phosphorus

Zhi-Xin Hu, [1, 2, §] Xianghua Kong, [1, 2, §] Jingsi Qiao, [1, 2, §], Bruce Normand[1,2] and Wei Ji[1,2,*]

[1]*Department of Physics, Renmin University of China, Beijing 100872, China*

[2]*Beijing Key Laboratory of Optoelectronic Functional Materials & Micro-Nano Devices,*
*Renmin University of China, Beijing 100872, China*

**ABSTRACT**

Stacking two-dimensional (2D) materials into multi-layers or heterostructures, known as van der Waals (vdW) epitaxy, is an essential degree of freedom for tuning their properties on demand. Few-layer black phosphorus (FLBP), a material with high potential for nano- and optoelectronics applications, appears to have interlayer couplings much stronger than graphene and other 2D systems. Indeed, these couplings call into question whether the stacking of FLBP can be governed only by vdW interactions, which is of crucial importance for epitaxy and property refinement. Here, we perform a theoretical investigation of the vibrational properties of FLBP, which reflect directly its interlayer coupling, by discussing six Raman-observable phonons, including three optical, one breathing, and two shear modes. With increasing sample thickness, we find anomalous redshifts of the frequencies for each optical mode but a blueshift for the armchair shear mode. Our calculations also show splitting of the phonon branches, due to anomalous surface phenomena, and strong phonon-phonon coupling. By computing uniaxial stress effects, inter-atomic force constants, and electron densities, we provide a compelling demonstration that these properties are the consequence of strong and highly directional interlayer interactions arising from electronic hybridization of the lone electron-pairs of FLBP, rather than from vdW interactions. This exceptional interlayer coupling mechanism controls the stacking stability of BP layers and thus opens a new avenue beyond vdW epitaxy for understanding the design of 2D heterostructures.

___________

[§] These authors contributed equally to this work.
[*] wji@ruc.edu.cn, http://sim.phys.ruc.edu.cn

## I. INTRODUCTION

Few-layer black phosphorus (FLBP), shown in Figs. 1(a)-(c), has recently been predicted theoretically and demonstrated experimentally to be an ideal material for next-generation nano-, thermo- and optoelectronics [1-5]. One unique property that distinguishes FLBP from all the other systems in this 2D materials revolution is its anisotropy: theoretical predictions [3,6,7] of geometric [3,7] and electronic structures [3,7] have been followed closely by experimental observations [1,5] of anisotropic electric (carrier mobility) [1,3,5,7], optical (light absorption) [3,5] and mechanical (elastic moduli) [3,6,8] properties. A less well recognized but equally striking difference of FLBP from other 2D materials lies in its strong interlayer coupling, which is mediated by interactions much stronger than van der Waals (vdW) [3]. Density functional theory calculations [3,7] identify direct wavefunction overlap between BP layers, leading to bonding- and anti-bonding-type features among the layers, which broaden both valence and conduction bands and give rise to a thickness-dependent bandgap [3]. Given these two distinctive properties of FLBP, we would also expect remarkable anisotropy and layer-dependence in its lattice dynamical (vibrational) behavior. Although highly relevant to mechanical, electronic and optical properties, the phonon spectrum of FLBP, and in particular its dependence on sample thickness, has yet to be characterized accurately.

The vibrational properties of multilayer 2D materials, most notably graphene [9-16] and transition-metal dichalcogenides (TMDCs) [17-26], have been studied intensively, in particular their interlayer optical modes. These have a much higher intensity than the interlayer acoustic (IA) modes and normally show a blueshift when the sample thickness (layer number) is increased [17]. They are also rather sensitive to changes in atomic environment, such as local strain or surface adsorption. We showed previously [3] that FLBP has a thickness-dependent variation of lattice constants, which implies an evolution of optical-mode frequencies. By contrast, IA modes reflect directly the strength of interlayer coupling in the directions perpendicular (breathing modes) and parallel (shear modes) to the layer plane of a 2D material. Recent observation of the rather weak IA modes in graphene [12,13,15,19-21] makes possible the investigation of interlayer coupling strengths in stacked pristine 2D materials and in heterostructures comprised of different 2D materials. Building such a heterostructure layer by layer is known as vdW epitaxy [27], because the vdW force dominates the interlayer coupling in graphene, BN and TMDCs. However, a full understanding of all types of interlayer interaction is required to control the growth of multilayer structures, the likely form of any practical device exploiting the properties of 2D materials [27-33], and here we reveal a different dominant interlayer coupling mechanism in BP.

In this article, we show by high-precision *ab-initio* calculations that the three Raman-observable optical modes,

i.e. $A_g^2$ (atoms moving along *x*, Fig. 1(d)), $B_{2g}$ (along *y*, Fig. 1(e)) and $A_g^1$ (along *z*, Fig. 1(f)), undergo an anomalous redshift with increasing layer thickness. Quite unexpectedly, it is the interlayer coupling which governs this trend, although changes of the lattice constant (strain effects) are partially responsible for its magnitude. Further, the $B_{2g}$ and $A_g^1$ modes split into two separate branches, corresponding to motion of atoms in the surface and inner layers, due to their differing effects on the hybridized electronic states. That the surface frequency is higher, suggesting stronger interlayer interactions at the surface, is absolutely new in multilayer 2D materials. The same result is found for the IA modes, other than the shear mode in the armchair direction (Sx, Fig. 1(g)), due to its coupling to the $B_{1u}^{2}{}'$ mode. The combined effects of strain and interlayer interactions (manifest as a phonon-phonon coupling) give rise to a blueshift of Sx, a redshift of Sy (shear in the zigzag direction, Fig. 1(h)) and a nearly unaffected Bz (breathing mode, Fig. 1(i)). Similar to other anisotropic properties [3], the explicit anisotropy of the shear modes in FLBP is understood by analyzing the electron hybridization through the distribution of (differential) charge density. We demonstrate that all of these striking results, so anomalous by comparison with graphene and TMDCs, originate from the direct interlayer wavefunction overlap of lone electron-pairs, a property of FLBP unique among 2D materials.

## II.  RESULTS

### A. Vibrational modes and dispersion relations in FLBP

Figures 1(a)-(c) show top and side views of the geometry of FLBP, together with the surface Brillouin Zone. In the primitive cell of FLBP, each layer contains four P atoms forming two sub-layers. Structural parameters *a* and *b* represent the lattice constants along the armchair and zigzag directions, *c* denotes the distance between the two sub-layers and *c'* the interlayer spacing. Bulk BP belongs to space group *Cmca* (No. 64) and to the $D_{2h}^{18}$ (mmm) point group containing 12 vibrational modes at the **G** point. With the symmetry reduced from 16 to 8 operators, FLBP is categorized into two different space groups, *Pmna* for odd and *Pbcm* for even numbers of layers, but it shares the same $D_{2h}$ point group with bulk BP and we label the modes accordingly. A full list of all 12 vibrational modes is available in Fig. S1. Three optical modes, $A_g^2$, $B_{2g}$ and $A_g^1$ (Figs. 1(d)-(f)), together with the three IA modes, Sx, Sy and Bz (Figs. 1(g)-(i)), are in principle Raman-observable and we focus on these six modes in the following discussion.

When two monolayers are stacked together, the phonon spectrum is doubled to include in- and out-of-phase partners for every mode. The splitting of these branches reflects the strength of interlayer coupling. Figures 1(j) and (k) show the phonon dispersion relations of mono- and bilayer BP, with the Raman-active optical modes of the monolayer and IA modes of the bilayer highlighted in color. For the in-phase mode $B_{1u}^2$ (Fig. 1(l)), a very strong splitting of 20.4 cm$^{-1}$ (relative to a mode energy of 127.0 cm$^{-1}$) is found from its partner $B_{1u}^2{}'$ mode (Fig. 1(m)) at the **G** point, while the 5.1 cm$^{-1}$ splitting of mode $B_{3g}^1$ is also appreciable. Modes $B_{1u}^2$, $B_{1u}^2{}'$ and $B_{3g}^1$ correspond respectively to shear modes along *z* (Figs. 1(l) and (m)) and a stretching one along *x* (Fig. 1(n)). They are correlated because stretching along *x* changes bond angle $\theta 1$ (Fig. 1(c)), which affects the intra-layer shear motion. The higher frequency of $B_{1u}^2{}'$ is easily understood from the spatial distribution of charge in the bonding state of the highest valence-electron wavefunction (VB1, shown in Fig. 2**g** of Ref. 3), which has significant interlayer electron density. We return below to a detailed discussion of this effect and of the mechanism for such strong interlayer coupling in FLBP.

We comment in addition that the frequency rise for Sx and Sy from the **G** point to the **X** or **Y** point is rather rapid, consistent with the fact that the intra-layer covalent bond is stiffer for stretching than for bending (represented by the dispersion of the Bz mode). The anisotropic character of FLBP is reflected clearly in the two IA shear modes, in that Sx is much softer than Sy. We will demonstrate below that the microscopic explanation of this anisotropy lies in the nature of the interlayer coupling, and not directly in the smaller elastic moduli of the *x* direction as reported in our previous study [3].

B.  **Layer-dependence of vibrational frequency**

In a conventional weakly coupled 2D material, weak reinforcement of the planar structure causes the frequency of optical phonons to rise with sample thickness, i.e. a blueshift [17]. In FLBP, the evolution of these phonon frequencies is anomalous, showing a decreasing trend (redshift) in all three cases (Figs. 2(a)-(c)), although $A_g^2$ does show a slight increase (Fig. 2(a)) beyond four layers (4L) and a splitting appears at 3L for $B_{2g}$ and $A_g^1$. This splitting effect is very weak (1 cm$^{-1}$) for $B_{2g}$, whose frequency decreases continuously (Fig. 2(b)), but for $A_g^1$, it reaches a maximum of 7 cm$^{-1}$ and one branch increases while the other reaches a minimum at 4L (Fig. 2(c)). The associated vibrational displacement pattern indicates that this frequency splitting, which is completely different from that in Fig. 1(k), arises from the quite different motion of surface and inner atoms. The surface atoms vibrate significantly more strongly than the inner atoms and contribute to the higher-frequency (upper) branch, whence we introduce the terminology "surface splitting." We defer an explanation of this counterintuitive effect (normally surface atoms have lower restoring forces and thus lower frequencies), and of the mysteriously anomalous layer-dependent redshift, until the presentation of results associated with Figs. 3 and 4.

The IA modes also show abnormal behavior with increasing layer thickness. An *N*-layer BP (NL-BP) system has *N-1* interlayer shear modes in each direction and *N-1* breathing modes. The uppermost branch, with the highest frequency, represents out-of-phase *z*-axis motion of all pairs of adjacent layers, while the lowest branch corresponds to the mode in which one half of the NL-BP sample oscillates collectively and out of phase with the other half (see Figs. S2 to S4). Frequencies for the three modes may be predicted for all *N* using a one-dimensional (chain) model where the nearest-neighbor force constant is calculated from bulk BP (see the Methods section). Calculations of the Sx, Sy and Bz frequencies for bulk BP give the values 24.7 cm$^{-1}$, 48.5 cm$^{-1}$ and 89.3 cm$^{-1}$, consistent with experimentally measured values of 19 cm$^{-1}$, 52 cm$^{-1}$ and 87 cm$^{-1}$, respectively [34-36]. The chain model then derives frequencies of 17.5 cm$^{-1}$, 34.3 cm$^{-1}$ and 63.1 cm$^{-1}$ in bilayer BP, and the expected fan-like feature of layer-dependent frequencies for increasing *N* (Davydov splitting), shown by open symbols with lighter colors in Figs. 2(d)-(f).

Our results from density functional perturbation theory (DFPT) for 2L- to 6L-BP, shown in Figs. 2(d)-(f) by the darker solid symbols, differ quite significantly from the chain model in some cases. Some Sx modes lie substantially lower and the frequency distribution is distorted away from the chain-model results (Fig. 2(d)), with the largest difference of 4.7 cm$^{-1}$ for 2L-BP reducing to 1.2±0.6 cm$^{-1}$ for 6L-BP. By contrast, for mode Sy, the calculated frequencies lie significantly closer to, but instead above, the chain-model values (Fig. 2(e)): the 2L-BP frequency is only 2.3 cm$^{-1}$ larger and the deviation is rather small for other layer numbers, with a crossover of chain and DFPT values for the lowest-frequency branch. Unlike the shear modes, for mode Bz the DFPT results are nearly identical to the simple expectations encoded in the chain model. These anomalous frequency variations, occurring for both optical and IA modes, have the same microscopic origin as the other properties that make BP quite distinct from other 2D materials, and we return to this topic below.

We summarize briefly the situation for experimental observation of these modes. NL-BP has a center of inversion symmetry for both odd and even *N*, and the Raman-active modes preserve this symmetry when vibrating. This selection rule applies to both shear and breathing modes and, as a result, all modes in the lowest branch, i.e. $Sx(y)_N^1$ and $Bz_N^1$, are Raman-active. The highest-branch modes ($Sx(y)_N^{N-1}$ and $Bz_N^{N-1}$), are Raman-active only for even *N*, the second-highest ($Sx(y)_N^{N-2}$ and $Bz_N^{N-2}$) for odd *N*, and so on (details are available in Figs. S2-S4). Thus Raman spectroscopy can provide experimental confirmation of the unconventional frequency shifts we predict for the optical and IA modes in FLBP.

Turning to the static properties of NL-BP, Figs. 2(g)-(i) show the lattice constants we compute for all values of *N*. The armchair lattice constant, *a*, decreases by 0.13 Å from 1L to 6L (Fig. 2(g)), whereas the zig-zag one, *b*, is slightly stretched, albeit by less than 0.02 Å (Fig. 2(h)). The vertical distances between two sub-layers (*c*) or two layers (*c'*) are both smaller at the surface than between the inner layers (Fig. 2(i)), suggesting a stronger interlayer attraction at the surface, which is completely novel by comparison with other 2D materials, including graphene and $MoS_2$. The surface contraction and changes in structural parameters, equivalent to an external strain, are key factors determining the surface splitting and the anomalous redshift.

## C. Stress effect on frequency shifts

To gain more insight into the factors affecting the phonon modes, we investigate strain effects by applying a uniaxial stress along the *x* direction. Figures 3(a)-(c) show the stress-induced structural deformation of monolayer and bilayer BP. The corresponding stress-induced frequency shifts of $A_g^2$, $B_{2g}$ and $A_g^1$, shown in Figs. 3(d)-(f), are qualitatively similar in that increasing uniaxial stress causes modes $A_g^2$ and $B_{2g}$ to soften, whereas $A_g^1$ hardens. These results, analyzed in detail in Fig. S5, are due primarily to the reduction of bond angles as the bond lengths are nearly unaffected by the application of stress [3].

The IA modes are also sensitive to external stress, with mode Sx showing the strongest variation (Fig. 3(g)) while that for Sy is very weak (Fig. 3(h)). Both Sx (Fig. 3(g)) and Bz (Fig. 3(i)) share the tendency that their frequency rises almost linearly with the applied stress, whereas the Sy mode frequency shows a non-monotonic stress dependence with a maximum at -0.15 GPa, behaving similarly to Sx and Bz below this value but oppositely up to 0.6 GPa (Fig. 3(h)). A better understanding of these stress-induced effects may be obtained by visualizing the interlayer interaction as two springs connecting the two layers. In Figs. 3(j)-(l), atoms and springs shown in lighter colors indicate the equilibrium situation and darker colors the situation under stress; the width of the spring indicates its strength, becoming more attractive at shorter distances. When *a* or *c'* is decreased, the interlayer attraction is enhanced (Figs. 3(j) and (l)), resulting in a blueshift of Sx or Bz. However, such a decrease is always accompanied by a slight stretching of *b*, which reduces the restoring force along *y* (Figs. 3(b) and (h)). Both the reduced force and the enhanced attraction compete to control the frequency of the Sy mode, resulting in its non-monotonic behavior. The effect of this uniaxial stress on the lattice constants (Figs. 3(a)-(c)) is qualitatively very similar to the effect of increasing layer thickness (Figs. 2(g)-(i)) and therefore these stress-dependent considerations may provide a partial explanation for the blueshift of Sx and redshift of Sy in progressively thicker samples. However, details including the minimum in Sy frequency at 4L and the insensitivity of Bz remain unclear.

Returning to the optical modes, stress effects cannot explain the surface splitting and the anomalous frequency shifts observed in Figs. 2(a)-(c). However, Figs. 3(a)-(f), with an applied external stress, allow us to deduce a relationship between the change in structural parameters and the frequency shift. By mapping this stress-induced frequency shift to the layer-dependent frequency shifts in FLBP (Figs. 2(a)-(c)), we use the corresponding lattice parameters (Figs. 2(g)-(i)) to separate out the direct effects on frequency from structural deformation. Solid symbols in Figs. 4(a)-(c) denote the intrinsic DFPT frequencies, while open symbols labeled ``Stress'' show the stress-induced frequency shift. The stress-induced contributions do reproduce the intrinsic shifts in part for $A_g^2$ and $B_{2g}$ (Figs. 4(a) and (b)), although the minimum at 3L for $A_g^2$ is missing and the redshift for $B_{2g}$ is significantly smaller than the intrinsic shift. For the $A_g^1$ mode, however, the stress-induced contribution is opposite to the intrinsic one. It is, of course, not realistic to expect that the frequency shifts could be reproduced by the stress contribution alone, as many other factors are involved, including notably the surface contraction, which is thought to depend on interlayer interactions. The results of this comparison do confirm that the known structural deformation is only one of the key factors governing the frequency shifts, leaving gaps in our explanations that will most likely be filled only by a full understanding of the interlayer coupling.

**D. Role of interlayer coupling in vibrational frequency shift**

In 2D materials such as multilayer graphene and h-BN, the primary interlayer coupling is mediated by vdW forces, which generally have only weak effects on intra-layer (optical) vibration modes. By contrast, FLBP has a rather special structure in that each atom forms P-P bonds with three atoms in the same layer, leaving a lone electron-pair oriented out of the layer. These pairs interact strongly with others from adjacent layers, changing the electron distribution both within and between layers, and thus creating the situation in the anisotropic BP structure where not only the interlayer (shear and breathing) modes but also the intra-layer optical phonons may be weakened or strengthened. However, the stacking of BP layers is always accompanied by changes in the lattice parameters, which complicates efforts to isolate the effects of interlayer coupling alone.

As a further step towards this goal, we introduce an approximation in which *a* and *b* are fixed (see the Methods section) and show the resulting FLBP optical-mode frequencies, denoted by ``Layer,'' in Figs. 4(a)-(c). At a qualitative level, this estimate of pure interlayer coupling effects is remarkably successful in reproducing all the features of the DFPT results, including the minimum of $A_g^2$, the splitting of $B_{2g}$ and $A_g^1$, and the correct *N*-dependence of $A_g^1$. In particular, interlayer coupling softens all the optical modes, indicating weakened inter-atomic bonding within each layer. Such a weakening can be ascribed, as we show below, to the charge transfer of the lone electron-pairs from intra-layer P-P covalent bonds to interlayer bonds. The interlayer contribution to frequency softening for $A_g^1$ (Fig. 4(c)) counteracts the increasing trend of the ``Stress'' curve and the combination gives values very close to the DFPT results. For the other two modes, the interlayer-coupling curves have the correct (DFPT) tendency and the effective stress shifts them into quantitative agreement with the absolute DFPT values.

The splitting of the ``Layer'' curves also confirms that a stronger interlayer attraction at the two surfaces of FLBP separates the surface-atom motion from that in the inner layers, giving higher phonon frequencies at the surface. The electronic hybridization of lone electron-pairs gives a vivid microscopic illustration of this surface-separation phenomenon. Figure 4(d) identifies the uppermost six valence bands, VB1 to VB6, in 6L-BP. The electron density distributions of the six corresponding eigenstates, hybridized wavefunctions spanning the six layers, show very explicitly (Fig. 4(e)) that VB1 and VB2, the lowest-lying pair, are two bonding states for inner (layers 2 to 5) and surface (layers 1-2 and 5-6) layers, respectively, while VB6 and VB5 are their associated anti-bonding states. States VB3 and VB4 are the bonding and anti-bonding states for the surfaces weakly bonded to a central bilayer. The two stronger bonding states, VB1 and VB2, have dramatically different interlayer couplings which divide the six layers into two categories, with clear separation of the dominant atomic motion as reflected in the surface splitting of the $B_{2g}$ and $A_g^1$ frequencies. These results indicate that electronic hybridization of the lone electron-pairs dominates the interlayer coupling and plays a paramount role in governing the vibrational properties of FLBP.

**E. Interlayer force constant**

A quantitative measure of the interlayer coupling can be obtained from the Inter-Layer Force Constants (ILFCs), whose calculation is discussed in the Methods section. One of the striking properties of FLBP is that the ILFCs of the surface layers for the *y* and *z* directions are 4-5 % larger than those for the inner layers, as illustrated for 6L-BP in Fig. 5(a). Such an enhanced surface attraction is consistent with the surface contraction we compute (interlayer separations of 3.087 Å at the surface and 3.096 Å inside, Fig. 5(a)) and hence with the surface splitting of the shear- and breathing-mode frequencies. The differential charge density between the surface and inner layers (Fig. 5(b)) confirms the enhanced electron density, meaning stronger bonding, of the surface layer. This result also accounts for the surface splitting of the optical modes ($B_{2g}$ and $A_g^1$), because the shorter interlayer distance also enhances intra-layer charge density at the surface. By contrast graphene, often considered the prototypical layered material, has the opposite behavior, with a larger interlayer distance and 7% smaller ILFC along *z* (Fig. 5(a)) at the surface of a multilayer sample. These results confirm again that the qualitatively new physics of FLBP can be traced to the hybridization of lone electron-pairs, which provides strong and anisotropic interlayer coupling.

Indeed, the highly directional nature of the lone pair causes the ILFC of the *y* direction to be more than three times that of *x*. Vibration-induced charge redistribution increases the total energy of the system, determining the stiffness of the corresponding mode. Figure 5(c) shows the charge distribution at the center plane of two inner layers (Fig. 5(a)), where three lobes of charge density residing below or above the three P atoms are oriented in the *y* direction, highlighted by the black rectangle. Charge deformation along *y* therefore incurs much stronger resistance than that along *x*, the higher electron density causing much stronger mutual repulsion among these lobes along *y* and thus leading to the factor-3 ILFC anisotropy in the *x*-*y* plane of FLBP.

**F. Phonon-phonon coupling**

Compared with the enlarged surface ILFCs for *y* and *z*, the result for *x* is anomalous in that the surface layers have smaller ILFCs than the inner ones (Fig. 5(a)), despite the smaller interlayer separation. In analyzing the phonon dispersion relations for bilayer BP (Fig. 1(k)), we found that modes $B_{1u}^2$ and $B_{1u}^2{}'$ are strongly affected by interlayer coupling. As illustrated in Fig. 5(d), one of the strongest effects on mode $B_{1u}^2{}'$ is through its interaction with Sx. These two vibrational modes are particularly sensitive to interlayer electronic hybridization, specifically in the region of space highlighted by the red and green rectangles in Fig. 5(e). The $B_{1u}^2{}'$ mode introduces a *z*-component in the atomic displacements of mode Sx, allowing it to lower its frequency by reducing the disruption of the strong interlayer electron density; simultaneously, this interaction raises and splits the frequency of $B_{1u}^2{}'$ for the different layers.

This IA-optical phonon-phonon interaction provides the microscopic explanation for two of the anomalous features in our results. Although it acts to soften mode Sx for all branches (of Fig. 2(d)), and thus to depress the ILFC values for Sx (Fig. 5(a)), the effect is strongest for the unconfined surface layer because of its stronger out-of-plane vibrational motion; in fact it acts to reverse the order of *x*-direction ILFC values compared to *y* and *z*, causing the strong *x*-*y* anisotropy of FLBP (Table SI) and suppressing the anomalous surface splitting. The phonon-phonon coupling is also responsible for the discrepancy between DFPT and chain-model frequencies computed for the multilayer Sx modes in Fig. 2(d). Increasing the thickness of the sample reduces the out-of-plane atomic displacements (see Fig. S6) and the net surface contribution to the mode frequencies, as a result of which the discrepancy decreases systematically. As in our calculations under uniaxial stress, a minor opposing tendency due to directional charge redistribution governs the small discrepancy in Sy frequencies (Fig. 2(e)). The clear role of interlayer coupling in mediating phonon-phonon interaction effects is a further piece of compelling evidence for its importance in determining the vibrational properties of FLBP.

**III. Discussion**

The redshift we find for mode $A_g^2$ (Fig. 2(a)) is consistent with a recent experiment [37] in which the frequency was measured as 471.3 cm$^{-1}$ in a monolayer, decreasing to 470.0 cm$^{-1}$ for a bilayer and to 467.7cm$^{-1}$ for the bulk value. For the other two optical modes, significant broadening of the Raman peaks made the question of a redshift, blueshift or non-shift inconclusive. Again from Raman measurements, it was found for 5L-BP that decreasing temperature causes a blueshift of all three optical modes [8]. Because a temperature reduction normally means a shortened lattice constant, such a shift may be taken to reflect the effects of a uniaxial stress, as studied in Fig. 3. However, in FLBP there are two parameters for the *z*-axis lattice constant, with the "softer" interlayer *c'* varying faster than *c* and dominating the sample thickness (Fig. 3(c)), thus vastly complicating any stress-temperature relationship. A thermal blueshift of $B_{2g}$ and $A_g^1$ is nevertheless fully consistent with the stress-lattice-frequency relationship contained in Figs. 3(b), (c), (e) and (f). However, for mode $A_g^2$, the notion that *a* should reduce with decreasing temperature appears to be contradicted by the stress-induced redshift we observe. We comment that changes of *a* lie almost exclusively in changes of bond angle, not of bond length [3,6] (see Fig. S5), whence *a* is not guaranteed to fall at lower temperature and we suggest that our results might imply a negative thermal expansion [38,39] of FLBP in the *x* direction.

The two main sources of interlayer interactions in FLBP are expected to be electronic hybridization of lone electron-pairs and vdW interactions. Regarding the hybridization effect, although the lone pairs are attached in principle to atoms in a single layer, the interlayer electron densities we obtain (Figs. 4(e), 5(b), 5(c), 5(e) and Fig. S7) are similar to those of interlayer covalent bonds. We thus introduce the term "quasi-covalent" bonding for the interlayer coupling effect of lone pairs. Regarding vdW interactions, it is instructive to compare the ILFCs of FLBP with multilayer graphene, which is a purely vdW stacked material. Measurements of the breathing mode, which directly reflects the interlayer attraction, give a value of 111 ×10$^{18}$ Nm$^{-3}$ for graphene [40] and compare well with our calculated values of 110.7×10$^{18}$ Nm$^{-3}$ and 118.1×10$^{18}$ Nm$^{-3}$ for the surface and inner layers. The structure of monolayer BP is much more corrugated than graphene, causing the atoms to be 10% closer (~3.1 Å) to their adjacent layers and indicating a stronger vdW attraction. At first glance, the ILFC values in Fig. 5(a) seem to contradict this expectation (left side of Table SI), but we caution that these figures are normalized to surface area. Highly corrugated FLBP has, approximately, only half as many atoms per unit area involved in nearest-neighbor contact when compared to graphene. Indeed the ILFC per effective atom (right side of Table SI) is approximately 7.5 Nm$^{-1}$ for the *z* direction in BP, which is 2.5 times the theoretical value in graphene (< 3.1 Nm$^{-1}$). By extracting force constants from the shear-mode frequencies, the ILFC of graphene on a per-atom basis is 0.46 Nm$^{-1}$, which is 2/3 of the value for the "softest"

Sx mode in BP, and around 1/6 of the value for Sy. This dramatic difference between shear modes reinforces again the notion of a quasi-covalent interlayer bonding subject to the highly directional nature of the BP layers.

The large ILFC of FLBP in the $z$ direction may explain why isolating monolayer BP from the bulk using scotch tape appears to be more difficult than graphene. As we show in Fig. 5(a), the $z$-direction ILFCs of multilayer graphene are smaller at the surface but larger for the inner layers, which makes the separation of graphene monolayers tend to occur at the surface. For FLBP, the reversed ILFC values make exfoliation prone to happen between two inner layers and detachment of a monolayer from a thick BP sample very unlikely. However, the $x$-direction ILFCs of FLBP are weakest for surface attachment (Fig. 5(a)), suggesting that an easier route to obtaining monolayer BP may be by shearing a thick sample along $x$ rather than stretching along $z$.

In summary, we have demonstrated that FLBP has strong interlayer interactions quite different from other 2D materials such as graphene and $MoS_2$. The quasi-covalent electronic hybridization of lone electron-pairs between BP layers results in anomalous and layer-dependent vibrational properties for both optical and IA phonons. These include the redshift of optical modes, the separation of surface and inner atomic motion, the enhanced surface attraction, the complex evolution of the different IA modes and the strong anisotropy of the shear modes. This hybridization also introduces an IA-optical phonon-phonon coupling, an entirely new phenomenon in 2D materials, which softens the low-frequency Sx mode and suggests a reduced thermal conductivity in the $x$ direction. Quasi-covalent interlayer coupling makes FLBP quite distinct from other 2D materials in terms of anisotropic and layer-dependent electronic and optical properties, although the importance of the lone electron-pair to the hybridization mechanism is yet to be better recognized. In this work, we have extended the understanding of the exceptional nature of FLBP to its vibrational and, in part, its thermal properties. FLBP represents a novel category of 2D materials with highly anisotropic properties, due in no small part to this strong and directional interlayer coupling, and its tunable electronic, optical and vibrational (thermal) properties make it widely viewed in the community as an ideal system for designing performance devices. The key step in device fabrication is full control over the epitaxial growth of layered materials. In this regard, the lone electron-pair may play a much more important role than was previously appreciated in epitaxy and property refinement. Because the resulting interlayer coupling is significantly stronger than vdW interactions, the full understanding we offer here should be indispensable for the design and fabrication of multilayer devices both from pristine 2D materials and as heterostructures.

**Note added in proof:** during the completion of this work, an experimental study of phonon modes in FLBP appeared

[41], where Raman-scattering measurements of both low- and high-frequency breathing modes find the former to be very sensitive to interlayer coupling and hence to layer number. However, because the number of layers in the samples is not known, a quantitative comparison with our predictions is precluded. Very recent first-principles calculations by these and other authors [42] confirm some of the features revealed in our work, including phonon anisotropies and the potential for strain engineering. Although both sets of authors recognize the importance of interlayer coupling to their results, neither one provides a detailed analysis either of its physical effects or of its microscopic origin.

**Figures and Tables**

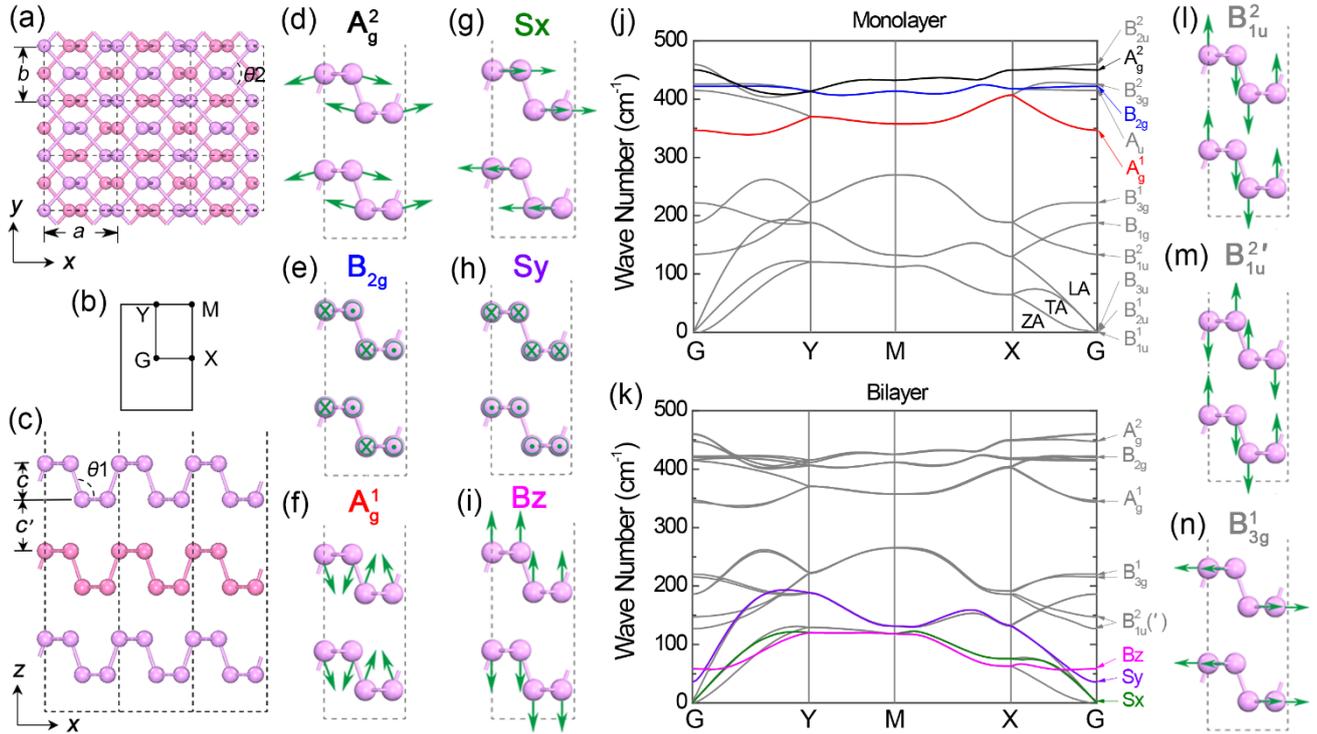

FIG. 1. Structure and phonon dispersions of FLBP. (a) Top view of 3L-BP, where *a* and *b* denote the lattice parameters in the *x* and *y* directions. (b) First Brillouin zone of a FLBP primitive cell, which was adopted in phonon dispersion calculations (panels j and k). (c) Side view of 3L-BP showing the definition of $c$, $c'$ and $\theta 1$. (d)-(i) Schematic diagrams for vibrational modes $A_g^2$, $B_{2g}$, $A_g^1$, shear x (Sx), shear y (Sy) and breathing (Bz) in 2L-BP. Green arrows indicate atomic displacements. (j,k) Phonon dispersion relations in monolayer (j) and bilayer BP (k) with mode labels at right; LA denotes the longitudinal acoustic, TA the transverse acoustic and ZA the out-of-plane acoustic mode of the monolayer. (l)-(n) Vibrational displacements of modes $B_{1u}^2$, $B_{1u}^{2\prime}$ and $B_{3g}^1$.

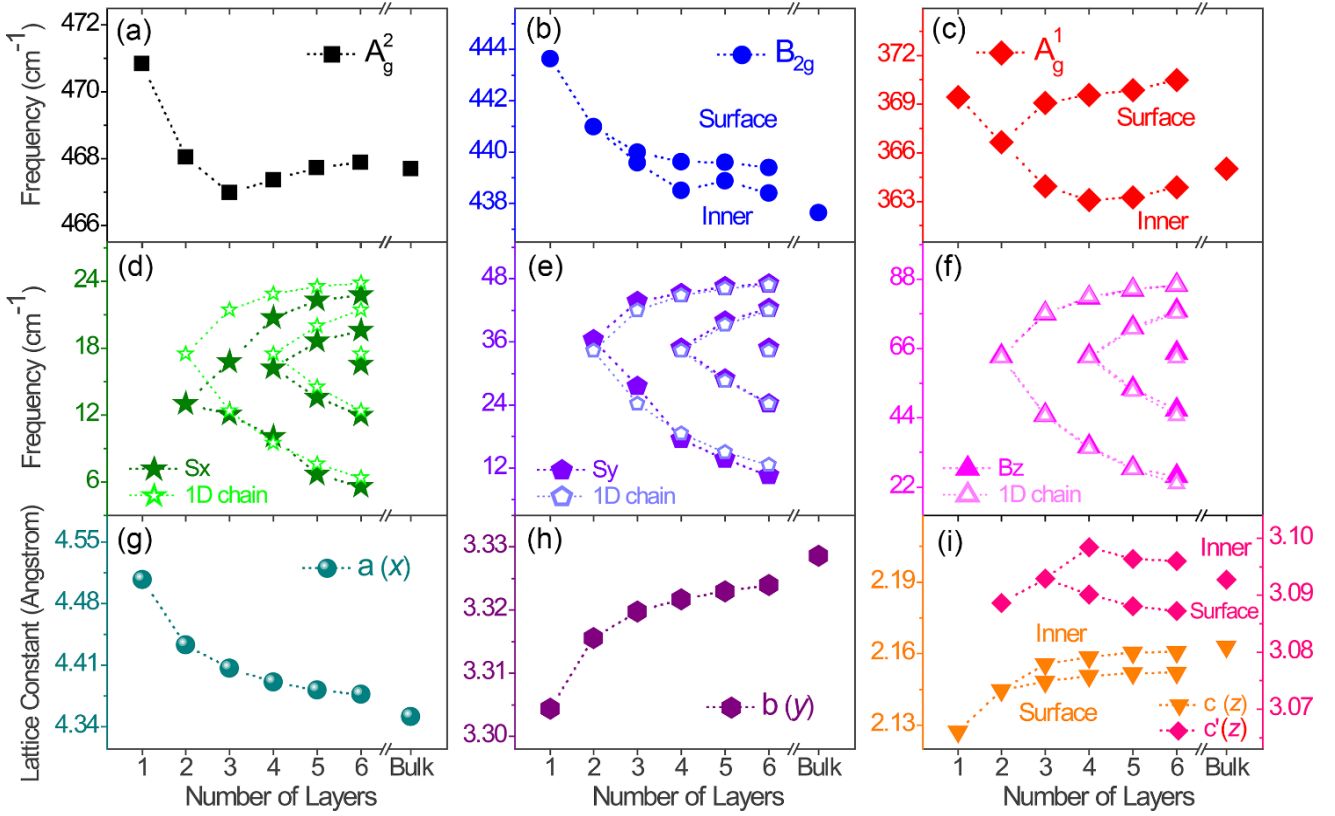

FIG. 2. Thickness-dependent vibrational frequency and lattice parameter. (a)-(c) Layer-dependent evolution of vibration frequencies at the G point for optical modes $A_g^2$ (a), $B_{2g}$ (b) and $A_g^1$ (c). In parenthesis after the mode label is the dominant direction of atomic displacement (Figs. 1(d)-(f)). The $B_{2g}$ and $A_g^1$ modes show a frequency-splitting between branches involving predominantly atoms on the surface layers ('Surface') and on the inner layers ('Inner') of the FLBP sample. (d)-(f) Layer-dependence of vibration frequencies at the G point for IA modes Sx (d), Sy (e) and Bz (f). Solid symbols show the calculated values and open symbols the results of a chain model using the computed bulk BP value as a reference. (g)-(i) Layer-dependence of lattice parameters $a$ (g), $b$ (h), $c$ and $c'$ (i), as defined in Fig. 1. Parameters $c$ and $c'$ also exhibit a surface-inner splitting (i).

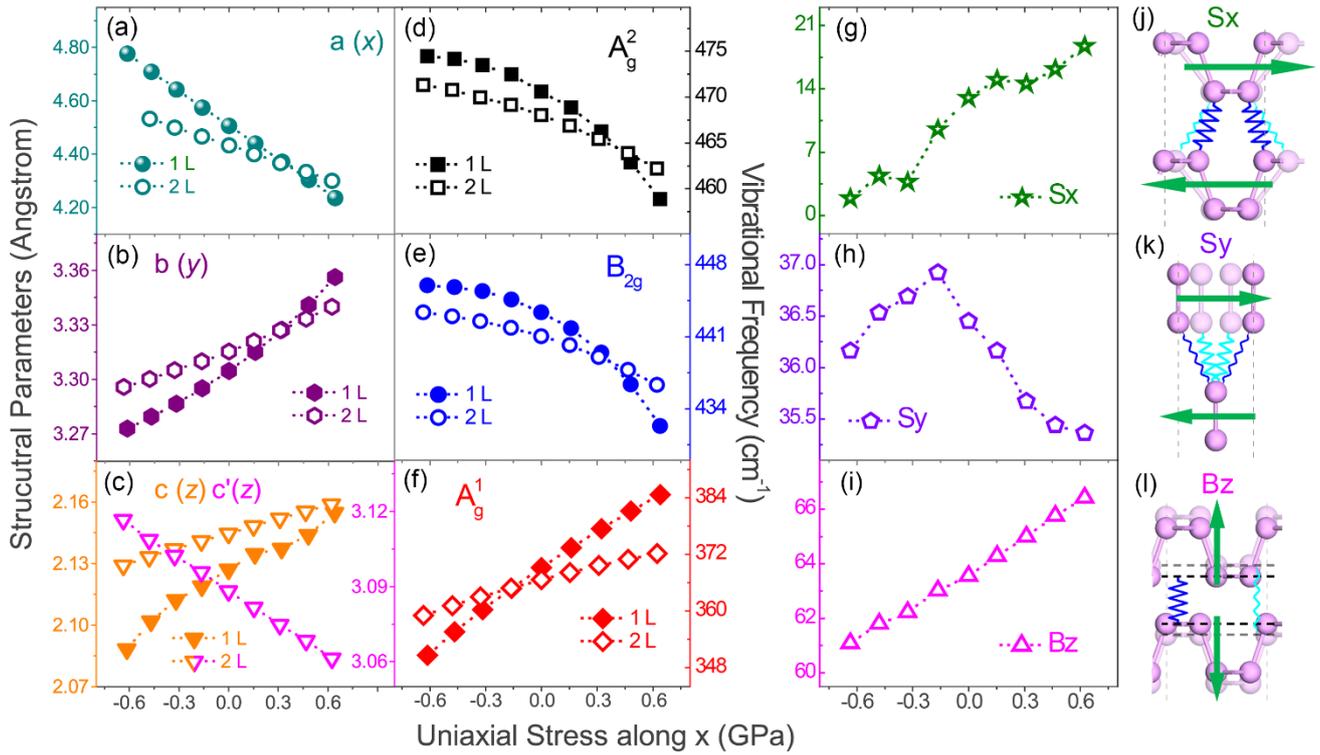

FIG. 3. Stress-dependence of vibrational frequencies. (a)-(c) Lattice parameters *a*, *b*, *c* and *c'* under a uniaxial stress along *x* for monolayer (solid symbols) and bilayer BP (open symbols). (d)-(i) Frequency response to external stress of modes $A_g^2$ (d), $B_{2g}$ (e) and $A_g^1$ (f) in monolayer and bilayer BP and modes Sx (g), Sy (h) and Bz (i) in bilayer BP. (j)-(l) Schematic illustration of variation in interlayer interactions under stress for Sx (j), Sy (k) and Bz (l). Background atoms with light blue springs indicate the structure and interlayer interaction strength without stress, while foreground atoms with dark blue springs show these under positive (compressive) stress along *x*. The deformation is exaggerated for clarity. Green arrows indicate the direction of vibrational motion in each mode.

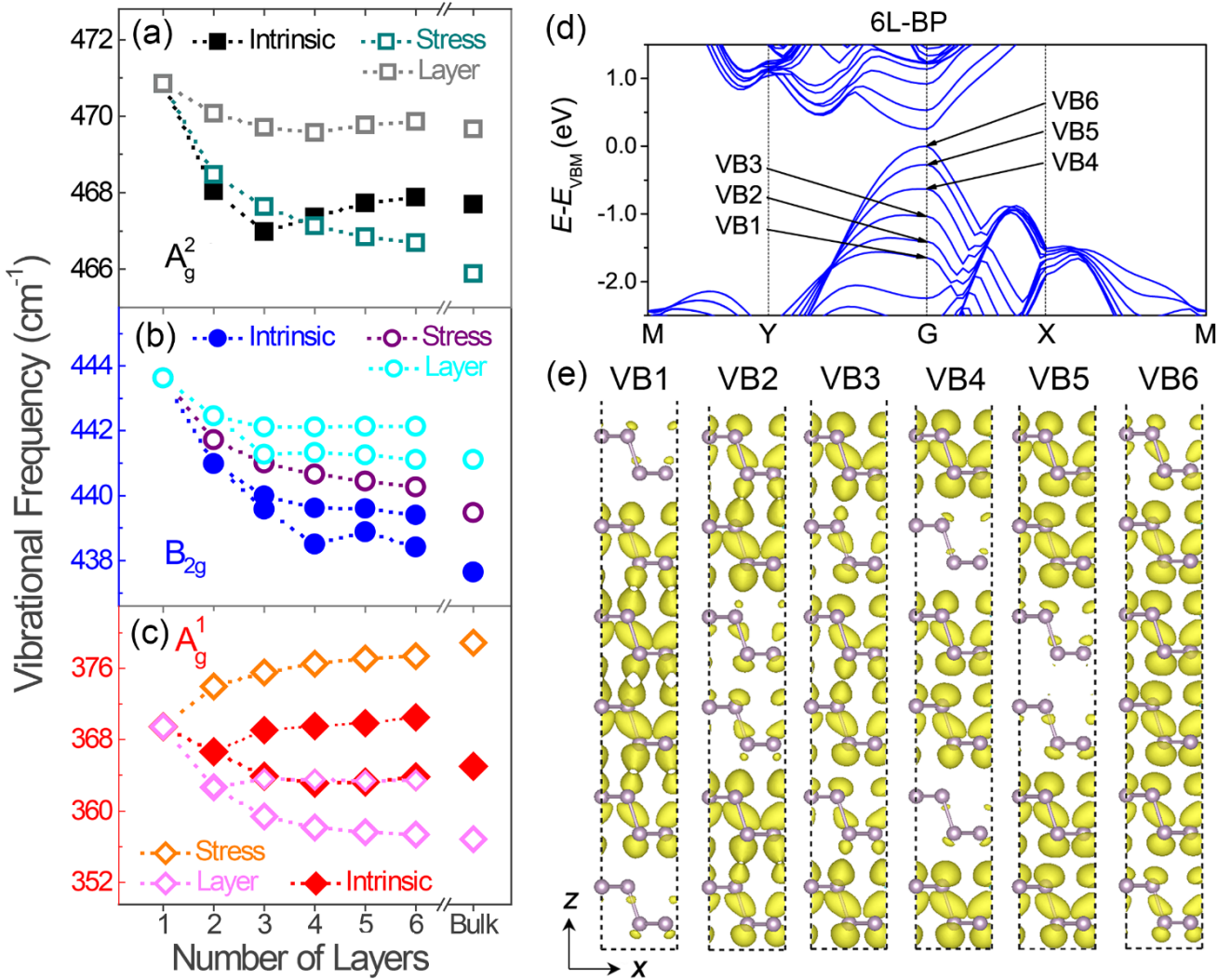

FIG. 4. Interlayer coupling and surface-inner separation. (a)-(c) Contributions from changes in lattice parameter and interlayer interaction to the thickness-dependent frequency. Curves ``Intrinsic'' (solid symbols) are the computed DFPT frequencies, identical to Figs. 2(a)-(c). Curves ``Stress'' are frequencies fitted using the stress-frequency and stress-lattice-parameter relationships obtained from Fig. 3. Curves ``Layer'' show only the contribution from interlayer coupling computed in the fixed-lattice approximation (see the Methods section). (d) Electronic bandstructure of 6L-BP with the six uppermost valence bands marked as VB1 to VB6 in ascending order. (e) Spatial distribution of valence-band wavefunctions VB1 to VB6, showing explicit surface-inner separation in the interlayer hybridization.

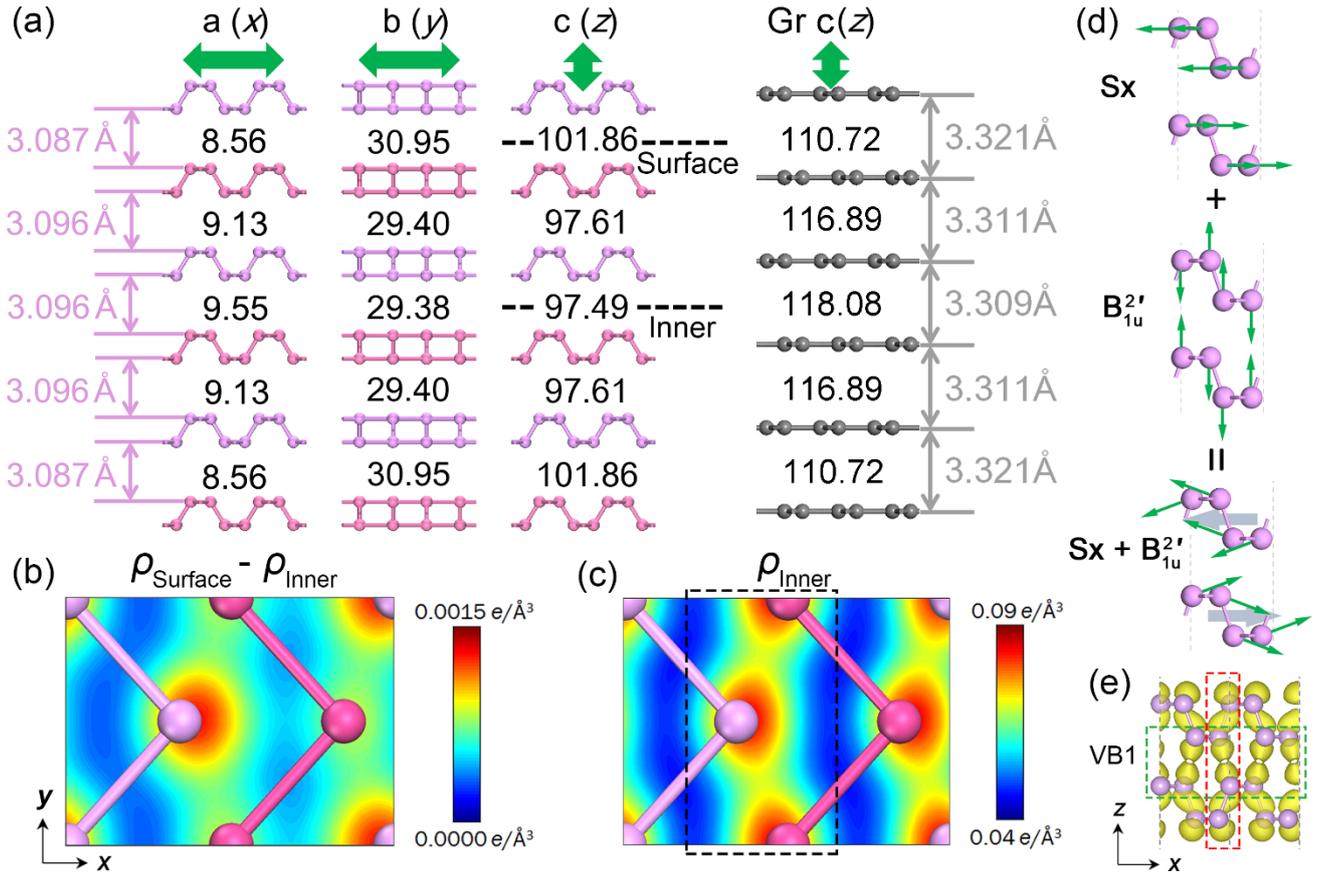

FIG. 5. Interlayer force constant and phonon-phonon interaction in FLBP. (a) ILFCs of each layer in 6L-BP along all three axes, compared with those of 6L-graphene ('Gr') for the c (z) direction, in units of $10^{18}$ Nm$^{-3}$. Interlayer distances are given on the left for BP and right for graphene. (b),(c) Differential charge density between a surface and an inner plane (b) and spatial distribution of charge density at the inner plane (c); the position of each interlayer plane is marked in (a). The black rectangle in (c) indicates a stronger electronic interaction along $y$, which is the origin of the stiffer Sy mode. (d) Illustration of the phonon-phonon coupling between Sx and $B_{1u}^{2}{'}$. Green arrows indicate the vibrational displacements of atoms and thick gray arrows of the whole layer. (e) Microscopic origin of phonon-phonon coupling: green and red rectangles highlight respective regions of the wavefunction affected by atomic motion in the Sx and $B_{1u}^{2}{'}$ modes. Strong electronic hybridization in the overlap region connects the motion of the two modes.

**ACKNOWLEDGEMENTS:**

**Author contributions:** W.J. conceived this research. X.K., J.Q. and W.J. performed layer-dependent structural relaxation and vibrational frequency calculations. X.K. and W.J. calculated phonon dispersions and vibrational frequencies for strained layers. Z.H. and W.J. carried out all calculations for rigid-layer modes. Z.H., J.Q., B.N. and W.J. analyzed the results and wrote the manuscript.

**Funding:** This work was supported by the National Natural Science Foundation of China (NSFC) under Grant Nos. 11274380, 91433103 and 11174365, the Ministry of Science and Technology (MOST) of China under Grant Nos. 2012CB932704 and 2012CB921704 and the Basic Research Funds of Renmin University of China from the Central Government under Grant No. 12XNLJ03. X.K. thanks the Chinese Scholarship Council for support. W.J. was supported by the Program for New Century Excellent Talents in Universities. Calculations were performed at the Physics Laboratory for High-Performance Computing of Renmin University of China and at the Shanghai Supercomputer Center.

**Competing financial interests:** the authors declare no competing financial interests.

## APPENDIX: METHODS

**1. Density functional theory calculation.**

Density functional theory calculations were performed using the generalized gradient approximation for the exchange-correlation potential, the projector augmented wave method [43,44] and a plane-wave basis set as implemented in the Vienna *ab-initio* simulation package (VASP) [45] and the Quantum Espresso (QE) [46] code. Density functional perturbation theory (DFPT) was employed to calculate phonon-related properties, including phonon dispersion relations (QE), Raman activity (QE) and shifts (VASP), vibrational frequencies at the Gamma point (VASP) and other vibration-related properties (VASP). The energy cut-off for the plane-wave basis was set to 700 eV for all calculations. A *k*-mesh of 31×31×1 was adopted to sample the first Brillouin zone of the conventional unit cell of FLBP in geometric optimization and phonon calculations. The mesh density of *k* points was kept fixed when calculating phonon frequencies for bulk BP using primitive cells. Fig. S8 shows the necessity of using such a dense *k*-mesh: coarser meshes give rise to imaginary values of the low-frequency optical mode $B_{1u}^2$ and inaccurate values for acoustic modes in systems thicker than 3L. In optimizing the system geometry, van der Waals interactions were considered at the vdW-DF [47,48] level with the optB86b (optB86b-vdW) [49,50] and optB88 (optB88-vdW) [49] exchange functionals. All results reported in the main text were calculated with the optB86b-vdW functional, which is thought to be more accurate in describing the structural properties of layered materials. [51] The shape (in-plane lattice parameters) of each supercell was optimized fully and all atoms in the supercell were allowed to relax until the residual force per atom was less than 0.0001 eV·Å$^{-1}$.

**2. Vibrational properties calculation.**

DFPT was used to compute the phonon dispersion (QE) and vibration frequency (VASP). The GGA functionals are known to underestimate the bandgap. Artificial overlap of valence and conduction bands is found in FLBP systems thicker than 3L, which leads to an imaginary frequency for the $B_{2u}^2$ mode. This issue was corrected largely by increasing the density of the *k*-mesh and also improved by imposing a breaking of structural symmetry; all results reported in the main text were obtained from a high-density *k*-mesh. Further discussion may be found in the Supplementary Material and in Fig. S8. For the optical modes, we followed the procedure [52] of applying a uniform scaling factor to obtain a clearer and more accurate comparison between theoretical and experimental phonon frequencies. In all our FLBP calculations we used a factor of 1.0587, which is the ratio of the experimental (467.1 cm$^{-1}$) [37,53] and computed (441.22 cm$^{-1}$) values for the $A_g^2$ mode frequency in bulk BP.

**3. Estimation of stress and layer effects on optical modes.**

The "Stress" curves in Figs. 4(a)-(c) represent frequency shifts induced by structural deformation for FLBP. The value of the frequency is defined as $\omega_{stress} = \frac{a - a_{1L}}{a_{1L}} p + \omega_{1L}$, where $a$ is the lattice parameter along $x$. The subscript "1L" denotes the value for 1L-BP. Variable $p$ is the gradient of frequency change in 1L-BP with $a$ close to $a_{1L}$, which can be derived from the values available in Figs. 3(a)**A** and (d)-(f). We used the exact value $a_{1L}$ of 1L-BP as the standard for its largest variation under stress; estimates using the other lattice parameters, $b$, $c$ and $c'$, as the reference do not change the results significantly. In FLBP, a variation in interlayer coupling is always accompanied by a change of lattice parameters, particularly $a$. As an approximate way to isolate the effects of interlayer coupling to lowest order, we fixed the lattice parameters $a$ and $b$ in all multilayer systems to those of 1L-BP, and the results of recalculating the phonon frequencies with these values are presented in the "Layer" curves in Figs. 4(a)-(c).

**4. Calculation of force constant.**

In an IA vibrational mode, the whole layer can be treated as one rigid body. The interlayer force constant (ILFC) $K$ is constructed by summing inter-atomic force constants over all atoms from each of the two adjacent layers, $K_{AB} = \sum_{a,b} D_{ab}$, $a \in$ [atoms in layer $A$], $b \in$ [atoms in layer $B$]. The matrix of inter-atomic force constants, essentially the Hessian matrix of the Born-Oppenheimer energy surface, is defined as the energetic response to a distortion of atomic geometry in DFPT [54] and takes the form $D_{ij} = \frac{\partial^2 E(\mathbf{R})}{\partial \mathbf{R}_i \partial \mathbf{R}_j}$, where $\mathbf{R}$ is the coordinate of each ion and $E(\mathbf{R})$ is the ground-state energy. The force on an individual ion can be thus expressed as $F_i = -\frac{\partial E(\mathbf{R})}{\partial \mathbf{R}_i}$. The vibrational frequencies $\omega$ are related to the eigenvalues of the Hessian matrix and the atomic masses by $\det \left| \frac{1}{4\pi^2 c^2} \frac{D_{ij}}{m_{ij}} - \omega^2 \right| = 0$, where $m_{ij} = \sqrt{M_i M_j}$ is the effective mass.

**5. Chain model.**

The 1D (chain) model proposed for the IA modes considers only the interlayer interaction between nearest neighbors and neglects any surface and substrate effects. The vibrational frequency is expressed as $\omega_{\text{1D-chain}} = \sqrt{\frac{k}{2\pi^2 c^2 m}(1 - \cos(\frac{(\alpha-1)\pi}{N}))}$, where $N$ is the number of layers and index $\alpha = 2, 3, 4, \ldots, N$.

# Interlayer electronic hybridization leads to exceptional thickness-dependent vibrational properties in few-layer black phosphorus

Zhi-Xin Hu,[1,2,§] Xianghua Kong,[1,2,§] Jingsi Qiao,[1,2,§], Bruce Normand[1,2],

Wei Ji[1,2,*]

[1]*Department of Physics, Renmin University of China, Beijing 100872, China*

[2]*Beijing Key Laboratory of Optoelectronic Functional Materials & Micro-Nano Devices, Renmin University of China, Beijing 100872, China*

## Supplementary Material

**Phonon modes.** Figs. S1-S4 provide complete information about the phonon modes in bulk BP and the acoustic modes in FLBP samples of two to six layers in thickness.

**Uniaxial stress.** Fig. S5 illustrates the origin and effects on the optical-mode frequencies of a uniaxial stress applied along the *x* axis.

**Phonon-phonon coupling.** Fig. S6 displays the differences between surface and inner branches of the low- and high-frequency Sx modes of FLBP from two to six layers.

**Interlayer electron densities.** Fig. S7 shows the results of a calculation of interlayer electronic density in 6L-BP, which illustrates clearly the strong interlayer hybridization arising from the lone-pair electrons of the P atoms.

___________
[§] These authors contributed equally to this work.
[*] wji@ruc.edu.cn, http://sim.phys.ruc.edu.cn

# I. DISCUSSION

**Imaginary frequency.** Figure S8 presents an example of the imaginary-frequency phonon obtained in FLBP systems thicker than 4L when the *k*-sampling is not dense enough. From its vibrational displacements, this mode results from the coupling of Sx and $B_{1u}^2$. A *k*-mesh of 13×13×1 was believed to be sufficiently fine for structural-relaxation and electronic-structure calculations, but in 4L-BP results in a mode at 26.13*i* cm$^{-1}$. The technical reason for this lies in the fact that standard DFT underestimates the bandgap, even when long-range correlation effects are included: FLBP is a narrow-gap semiconductor, but samples thicker than 3L are predicted to be metallic when a coarse *k*-mesh is used. We found that a denser mesh of 31×31×1 allows us to eliminate this imaginary frequency. Calculations performed by other groups appear not to have noticed this issue and may therefore be introducing an artificial structural instability.

The imaginary-frequency issue may also be solved by such an artificial breaking of structural symmetry along the vibration direction of the $B_{1u}^2$ mode, as shown in Fig. S8(b). Coarse-mesh calculations predict a metallic state for symmetric FLBP, whereas for asymmetric (broken-symmetry) FLBP a bandgap of 0.2 eV opens slightly off the Gamma point, as shown in Fig. S8(d). We have found that the DFPT phonon frequencies obtained with a denser *k*-mesh are rather reliable by comparing them with those of the broken-symmetry structure.

If a compressive stress is applied along the *z* direction, the valence and conduction bands overlap and the system becomes metallic. In this case, the inversion symmetry is genuinely broken, resulting in the structural deformation shown in Fig. S8(b). In addition, it could be inferred that low-energy electronic excitations may also lead to such a structural deformation through electron-phonon and phonon-phonon coupling, which would be a topic worthy of more detailed investigations.

## II. SUPPLEMENTARY FIGURES

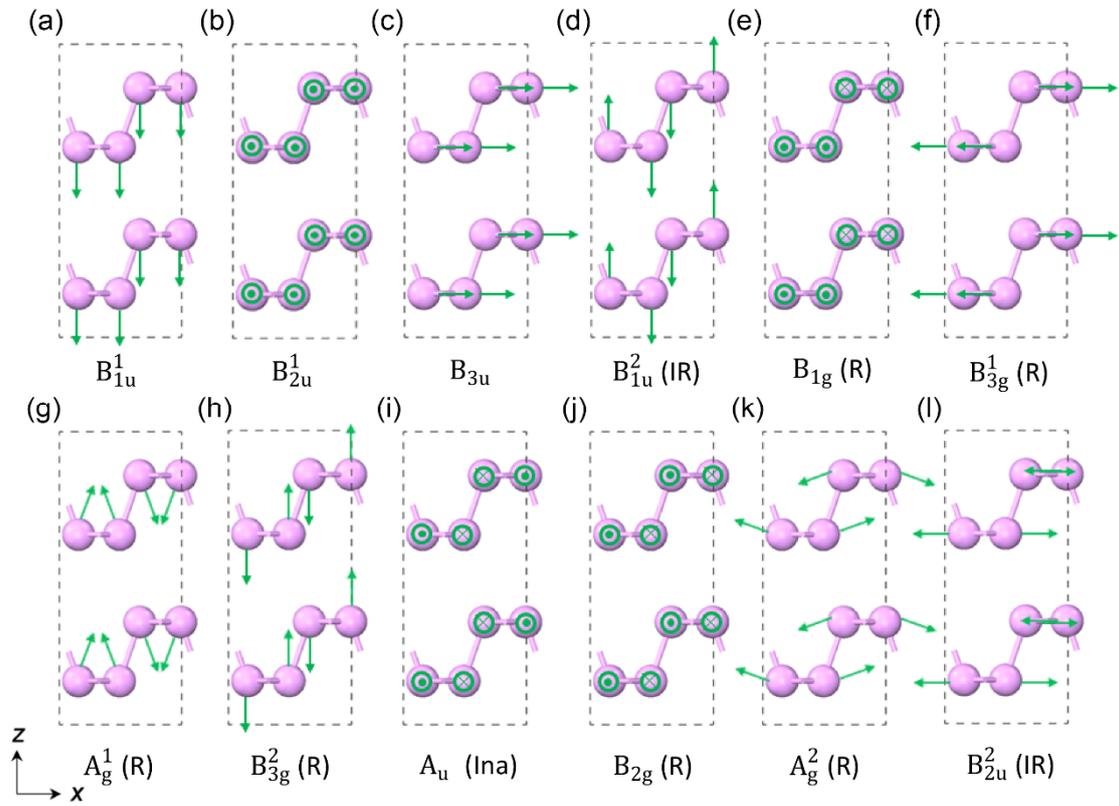

**Fig. S1. Vibrational displacements in bulk BP.** The atomic displacements, at the Gamma point in the conventional cell, are shown for all phonon modes. (a)-(c) Acoustic modes. (d)-(l) Optical modes. Labels ``R'', ``IR'' and ``Ina'' indicate respectively Raman activity, infrared activity and optical inactivity.

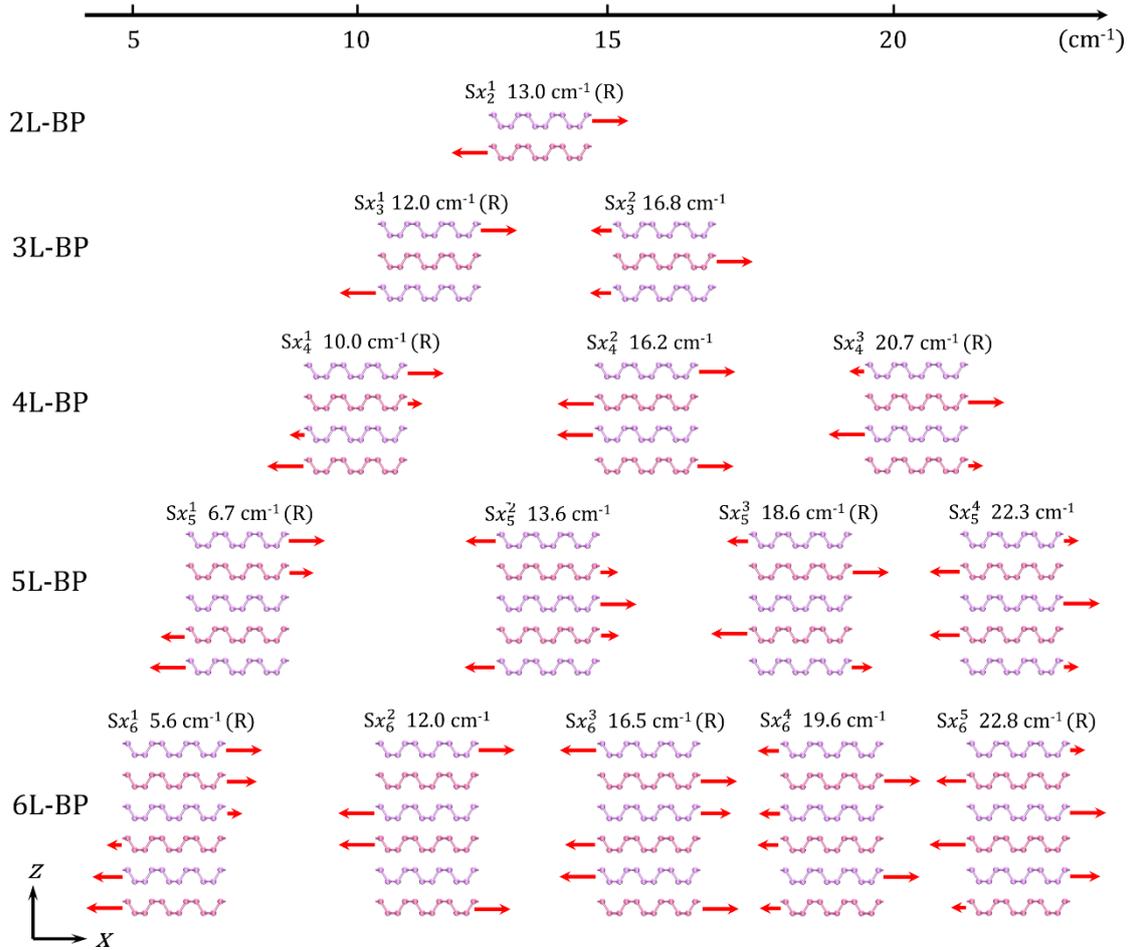

**Fig. S2. Vibrational displacements, frequencies and Raman activity of Sx modes.** Details of the Sx modes of FLBP, illustrating atomic displacements in all modes from two to six layers. Sx modes are present only in systems of two or more layers. Frequency values were calculated using the optB86b-vdw functional. Red arrows indicate the directions of displacement for each layer along the $x$ axis and arrow lengths represent the magnitude of vibrational motion. In the notation $Sx_N^n$ for each mode, subscript $N$ denotes the number of layers and superscript $n$, with values from 1 to $N$-1, identifies the branch in order of its vibrational energy from low to high. Label ``R'' indicates that the mode is Raman-active.

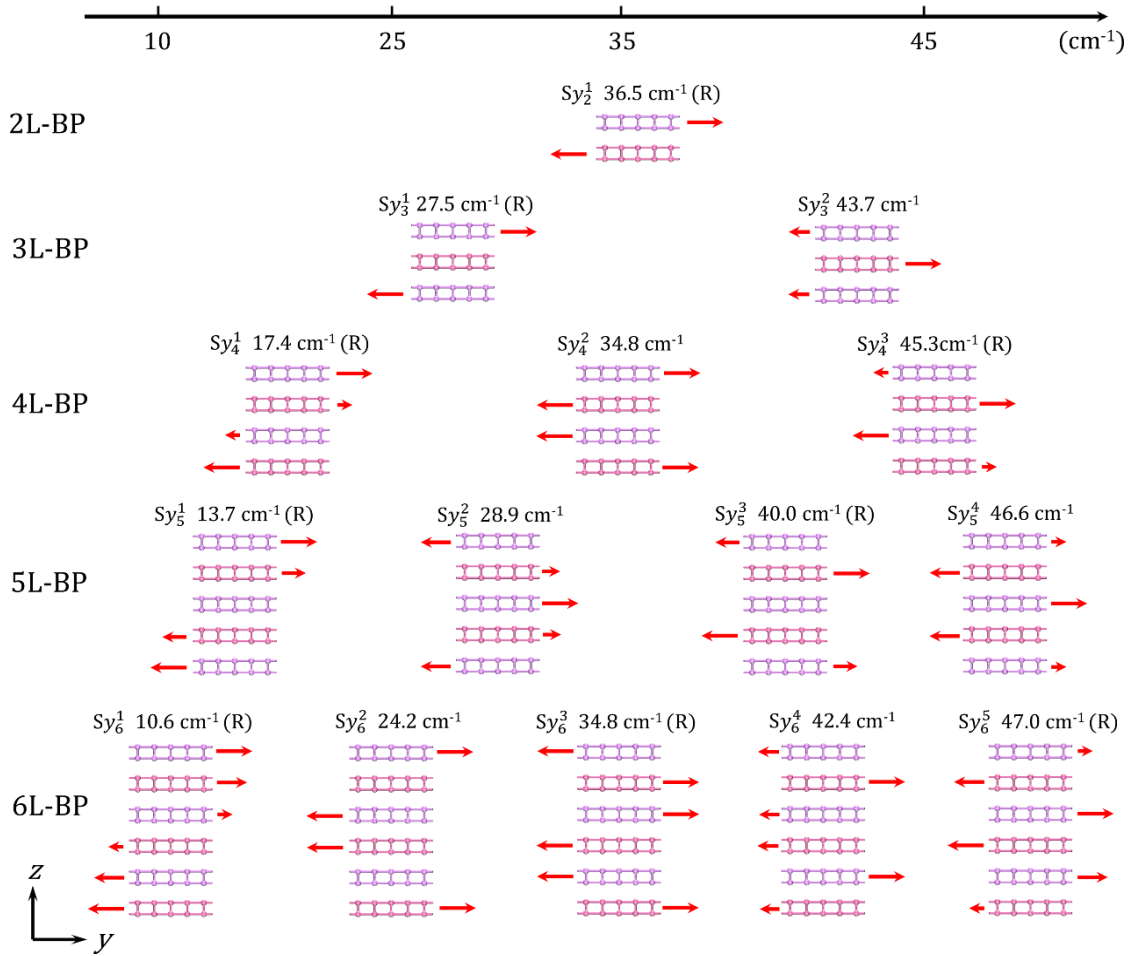

**Fig. S3. Vibrational displacements, frequencies and Raman activity of Sy modes**. Sy modes of FLBP from two to six layers; all notation (red arrows, $N$, $n$, ``R'') and the functional used are as in Fig. S2.

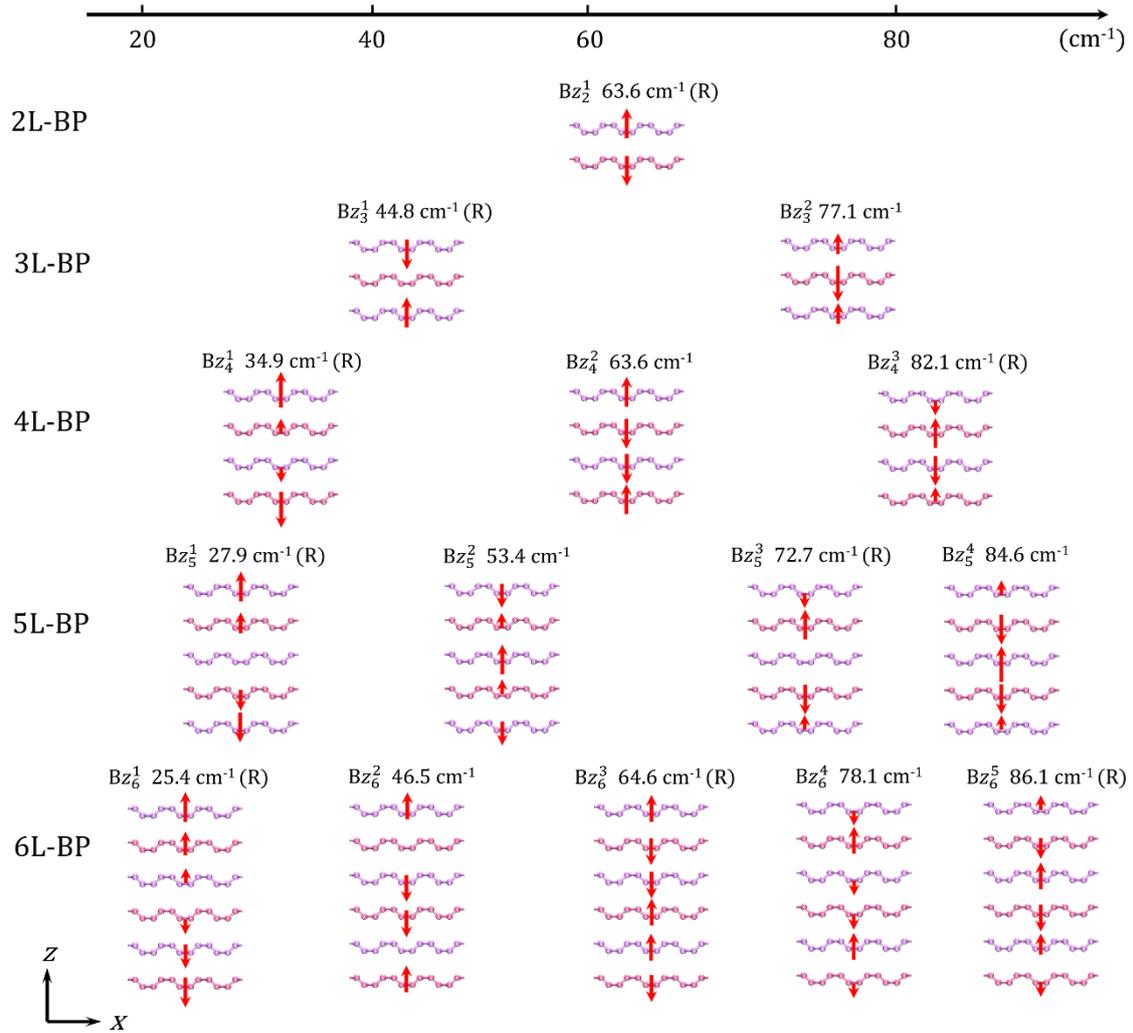

**Fig. S4. Vibrational displacements, frequencies and Raman activity of Bz modes.** Bz modes of FLBP from two to six layers; all notation (red arrows, $N$, $n$, ``R'') and the functional used are as in Fig. S2. The only difference is that these are breathing and not shear modes.

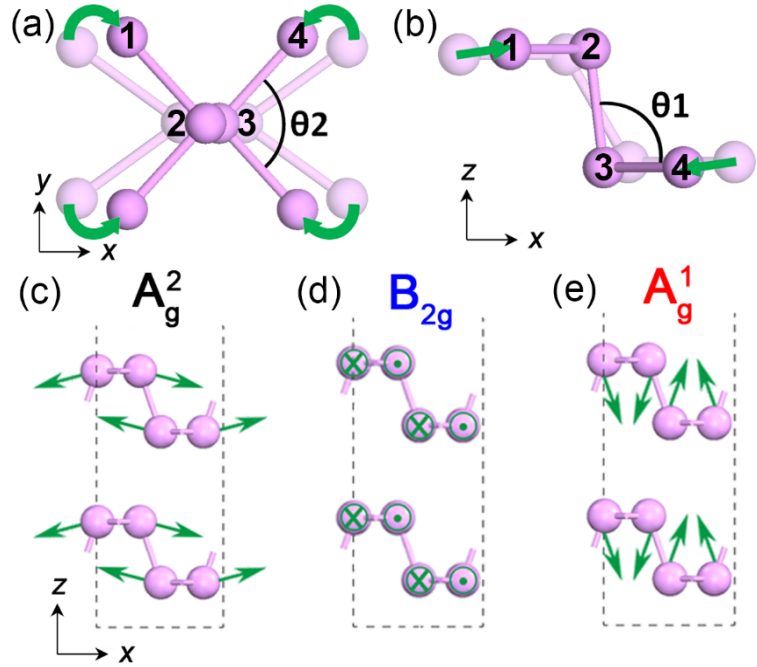

**Fig. S5. Stress effect on optical modes.** (a) Side and (b) top views of monolayer BP showing the structural deformation under uniaxial stress applied along $x$. Atoms in lighter colours in the background show the structure without stress, while foreground atoms in darker colours indicate the positions of atoms under positive (compressive) stress. Compressive stress along $x$ causes no appreciable change in P-P bond length, but bond angles do vary in that $\theta 1$ is reduced and $\theta 2$ enlarged. These bond angles modify the restoring force and moments during the vibrational motion. (c)-(e) Schematic illustration of vibrational displacements for modes $A_g^2$, $B_{2g}$ and $A_g^1$, respectively. For mode $A_g^2$ (c), atoms oscillate mostly along $x$ with a small $z$ component. The enlarged $\theta 2$ extends the $y$-axis separation of atoms 1 (3) and 2 (4), thereby lowering the restoring force along $x$ and causing a redshift under compressive stress. The behaviour of the $B_{2g}$ mode (d), where atoms 1 (3) and 2 (4) move in opposite directions along $y$, is similar. For the $A_g^1$ mode (e), the atomic displacements are largely in the same direction. The reduced angle $\theta 1$ causes the orientation of the bond connecting atoms 2 and 3 to be more parallel to the $z$ direction, enhancing the $z$-axis restoring force and creating a blueshift under compressive stress.

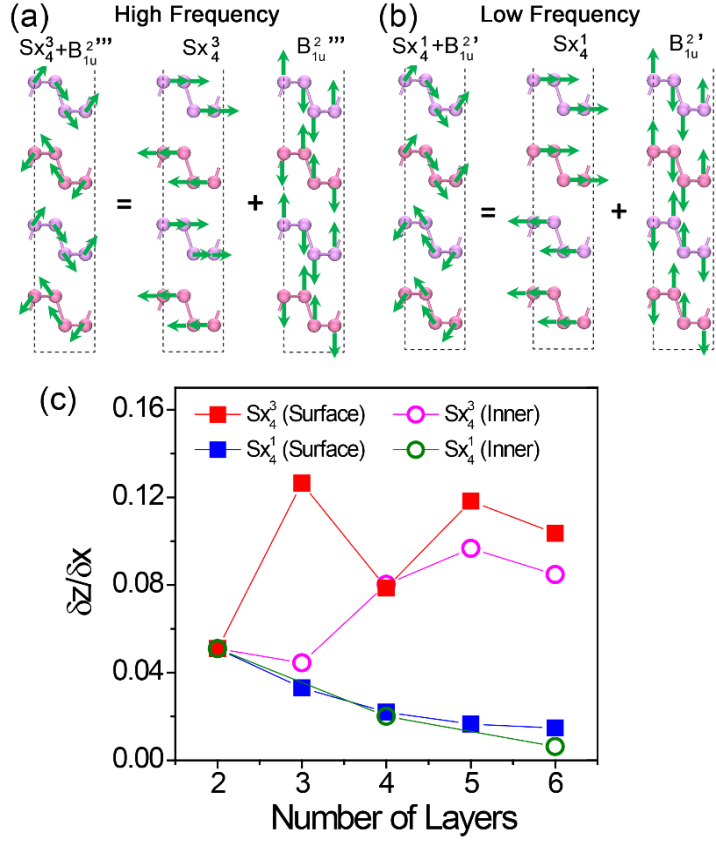

**Fig. S6. Spatial dependence of phonon-phonon coupling.** (a), (b) Coupling of highest- ($Sx_4^3$) and lowest-frequency ($Sx_4^1$) Sx modes with two $B_{1u}^2$ modes in 4L-BP. The coupled Sx mode has both *x* and *z* components, representing respectively in-plane shear and out-of-plane motion. The ratio of the two components, $\delta z/\delta x$, indicates the relative strength of out-of-plane vibrations. (c) $\delta z/\delta x$ for both highest- and lowest-frequency branches at the surface and inner layers. In the low-frequency branch, the central layers move only for systems with even layer number and thus are shown only for 2L-, 4L- and 6L-BP. The atomic displacements of the surface layers are always larger than those of inner layers, accounting for the ILFC values in Fig. 5(a). We find that this displacement drops with increasing layer thickness beyond 5L, consistent with the decreasing importance of surface contributions. Large differences in the $\delta z/\delta x$ ratio between the surface and inner branches in the highest-frequency mode explain the discrepancy between the DFPT and chain-model Sx frequencies for this branch, whereas the lowest-frequency branch shows little difference (Fig. 2(d)).

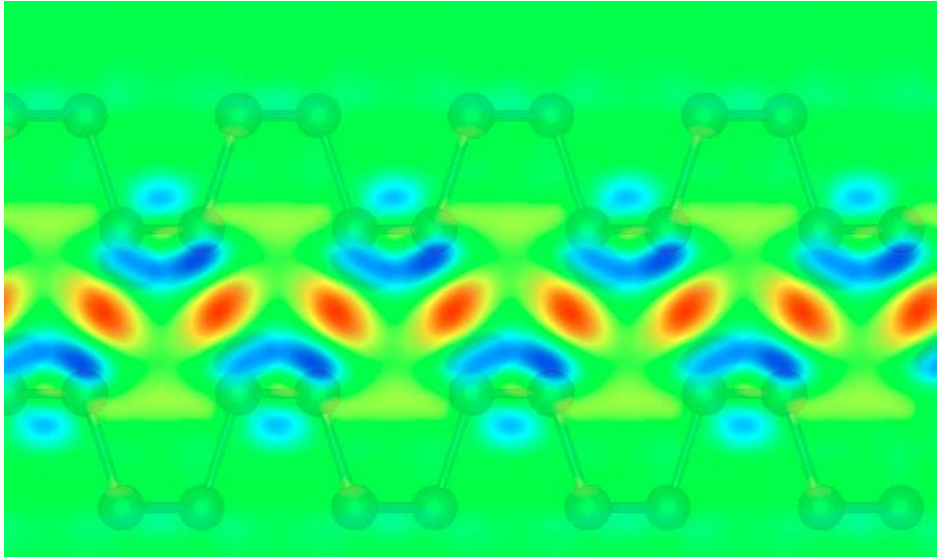

**Fig. S7. Differential charge density in FLBP.** Electronic density at the centre layers of 6L-BP, derived by computing $\rho_{6L\text{-}BP} - \rho_{3L\text{-}BP\_u} - \rho_{3L\text{-}BP\_d}$, where $\rho_{6L\text{-}BP}$ is the total charge density and $\rho_{3L\text{-}BP\_u}$ and $\rho_{3L\text{-}BP\_d}$ represent respectively the total charge density of the three-layer upper and lower halves of the system. Hot colours indicate charge accumulation and cold colours charge reduction. The clear reduction in electron density around the P atoms and accumulation in the region between the two layers are an explicit indication of ``covalent'' characteristics in the interlayer bonding. Because the interlayer electrons are from lone-pairs on a single atom rather than shared between atoms, we name this type of electronic hybridization ``quasi-covalent BP bonding''.

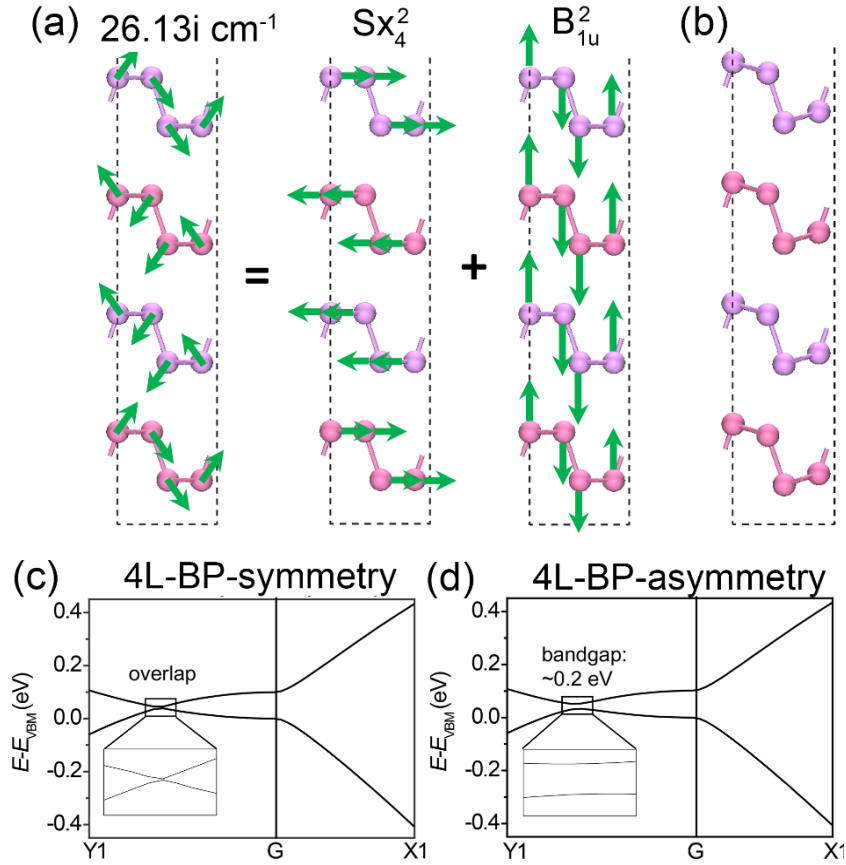

**Fig. S8. Source of imaginary frequency with coarse *k*-mesh.** (a) In 4L-BP, an imaginary phonon frequency appears at 26.13*i* cm$^{-1}$ when using a coarse *k*-mesh to sample the Brillouin zone. Decomposition of the vibrational displacements indicates that this mode consists of $Sx_4^2$ and $B_{1u}^2$, the two categories of phonon identified to have strong phonon-phonon coupling. (b) Structure of asymmetric 4L-BP obtained by a symmetry-breaking resulting from moving atoms along the vibration directions of $B_{1u}^2$. (c), (d) Electronic bandstructures for symmetric and asymmetric 4L-BP. The symmetric structure should be a small-gap semiconductor, but by standard DFT the valence and conduction bands are found to cross close to the Gamma point, as shown in the inset of panel (c). This band-crossing makes the Fermi surface unstable and results in a symmetry-breaking accompanied by the opening of a gap, as shown in panel (d). X1 is the point (0.1 0 0) and Y1 is (0 0.1 0).

## III. SUPPLEMENTARY TABLE

**Table SI | Interlayer force constants for FLBP and multilayer graphene.**

| | Per-area basis ($\times 10^{18}$ Nm$^{-3}$) | | | | Per-atom basis (Nm$^{-1}$) | | | | |
|---|---|---|---|---|---|---|---|---|---|
| Layer | Sx | Sy | Bz | Gr-Bz | Sx | Sy | Bz | Gr-S | Gr-Bz |
| 1-2 | 8.56 | 30.94 | 101.86 | 110.72 | 0.70 | 2.55 | 8.39 | | 2.91 |
| 2-3 | 9.13 | 29.40 | 97.60 | 116.89 | 0.75 | 2.42 | 8.04 | | 3.07 |
| 3-4 | 9.55 | 29.37 | 97.50 | 118.08 | 0.79 | 2.42 | 8.03 | 0.45* | 3.10 |
| 4-5 | 9.13 | 29.40 | 97.60 | 116.89 | 0.75 | 2.42 | 8.04 | | 3.07 |
| 5-6 | 8.56 | 30.94 | 101.86 | 110.72 | 0.70 | 2.55 | 8.39 | | 2.91 |

Calculated interlayer force constants for modes Sx, Sy and Bz of FLBP and Bz of multilayer graphene. The Bz force constants of FLBP appear smaller than those of graphene due only to the corrugated layer structure of FLBP; renormalized on a per-atom basis, even the values of the ``softest" (Sx) constants of FLBP are larger than those for the shear mode of graphene, while the values for Bz of FLBP are 2.5 times those of the corresponding graphene Bz modes.

* 0.45 Nm$^{-1}$ is the numerical average over five values varying from 0.39 Nm$^{-1}$ to 0.54 Nm$^{-1}$.